\documentclass{juliacon}
\usepackage[absolute]{textpos}
\usepackage[nolist]{acronym}

\begin{acronym}[Bash]
	\acro{SDN}{Software-Defined Networking}
    \acro{BGP}{Border Gateway Protocol}
    \acro{MD}{multi-domain}
    \acro{NBI}{Northbound Interface}
    \acro{IBN}{Intent-Based Networking}
	\acro{DAG}{Directed Acyclic Graph}
	\acro{MSA}{Multi-Source Agreement}
	\acro{OXC}{Optical Cross-Connect}
    \acro{API}{Application Programming Interface}
    \acro{KSPFF}{k-Shortest Paths First-Fit}
    \acro{RMSA}{Routing, Modulation, and Spectrum Assignment}
\end{acronym}

\begin{document}

% **************GENERATED FILE, DO NOT EDIT**************

\title{MINDFul.jl: A Framework for Intent-driven Multi-Domain Network coordination}

\author[1]{Filippos Christou}
\affil[1]{Institute of Communication Networks and Computer Engineering (IKR), University of Stuttgart}


\hypersetup{
pdftitle = {MINDFul.jl: A Framework for Intent-driven Multi-Domain Network coordination},
pdfsubject = {JuliaCon 2023 Proceedings},
pdfauthor = {Filippos Christou},
pdfkeywords = {Julia, Intent-based Networking, multi-domain, IP-optical},
}

\maketitle

\begin{textblock}{13.5}(3.4,1.6)
    {\color{blue}
    This work was presented in JuliaCon 2023 https://pretalx.com/juliacon2023/talk/review/YR83GQLFDC37NXFF7GFRV7LKGXHCVRWT 
    }
\end{textblock}

\begin{abstract}
    Network coordination across multiple domains is a complex task that requires seamless communication among network entities.
    Network operators aim to minimize costs while ensuring the requirements of user requests are met.
    Such efforts are highly challenging in decentralized environments with diverse network operators, where only partial knowledge of the complete network is available.
    Intent-driven multi-domain coordination offers various benefits, some inherent to \ac{IBN} and others stemming from the standardization of the \ac{NBI}.
    As standardization is still missing, there has not been a substantial initiative to develop tools that leverage this paradigm.
    \verb|MINDFul.jl| is a Julia library that provides a means to accelerate research in this area, at both the architectural and algorithmic levels.
    It provides a stateful, modular representation of common metro/core IP-optical network equipment as well as the common intent operations.
    Finally, it introduces a novel modular \ac{IBN}-over-\ac{SDN} architecture and is part of a library ecosystem that facilitates event-based simulations with a hackable interface and offers visualization support.
\end{abstract}

\acresetall

\section{Background}
\begin{figure}[t]
    \centerline{\includegraphics[width=5cm]{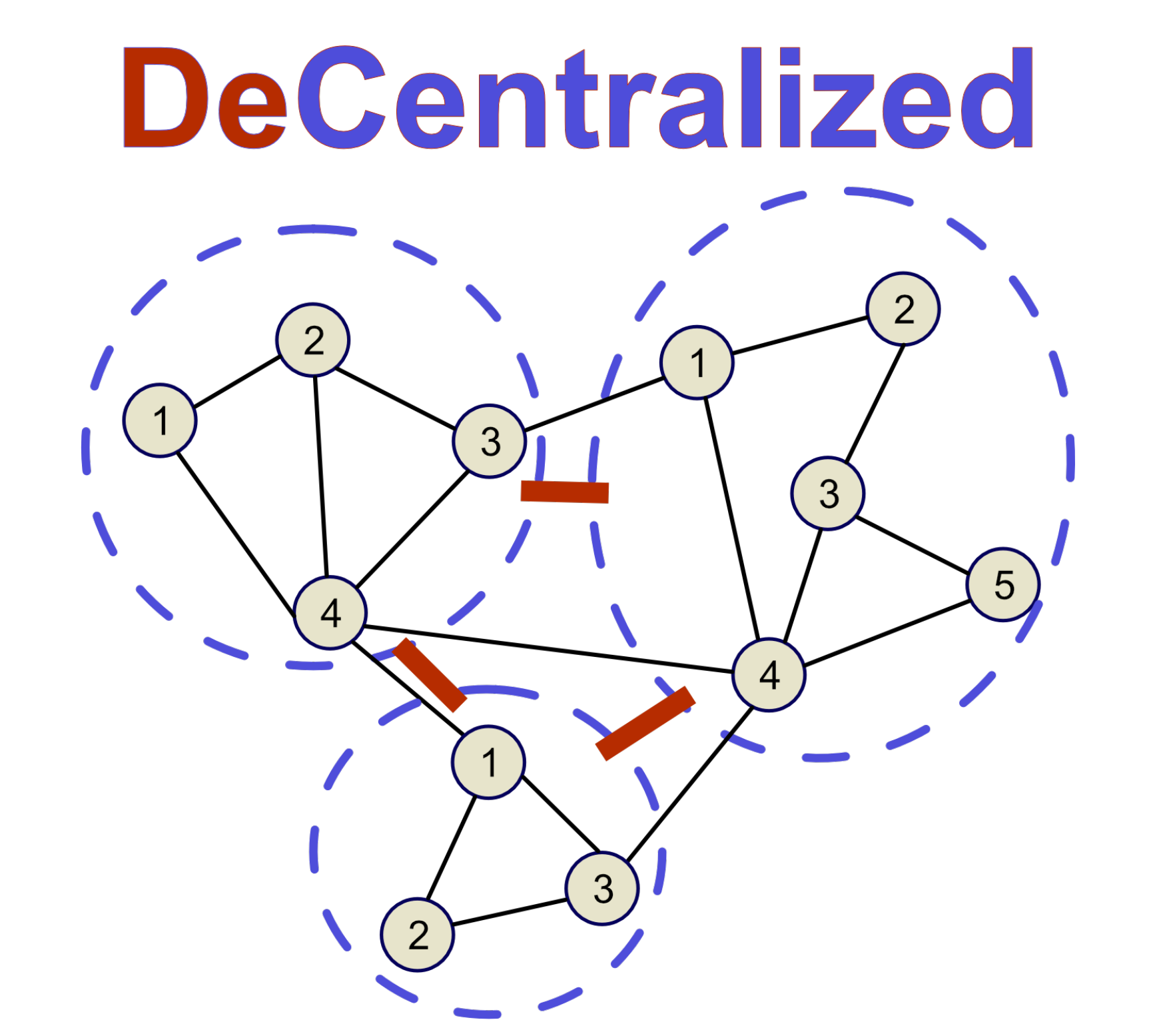}}
    \caption{Domains coordinate with each other in a decentralized fashion, while each domain has centralized control internally.}
        \label{fig:DeCentralized}
    \end{figure}

    As \ac{SDN} becomes more popular and many networks shift to centralized control for easier management and greater efficiency, \ac{MD} networking often must remain decentralized by its very nature.
    This will cause most of the networks to operate using a centralized \ac{SDN} controller internally, but still need to coordinate in a decentralized fashion with the neighboring domains, as shown in Figure~\ref{fig:DeCentralized}.

    An intent-driven approach \cite{2023ietf, 2022ChristouCNSM} has been proposed that can replace traditional \ac{MD} decentralized protocols like \ac{BGP}, since it offers greater flexibility in interactions and support for much wider network capabilities.
    \ac{IBN} provides a layer of abstraction where high-level objectives (i.e., intents) can be defined and automatically handled by the system.
    Several design and algorithmic decisions need to be made to develop an \ac{IBN} framework, including defining an intent state machine and the algorithms that enable state transitions.
    Commonly, an intent has at least the following four states, although naming conventions might differ:
    \begin{itemize}
        \item \emph{uncompiled} for unprocessed intents inside the system
        \item \emph{compiled} for processed intents with a well-defined implementation
        \item \emph{installed} for active intents whose implementation has been realized in the appropriate network devices
        \item \emph{failed} for active intents that malfunction after an operation failure.
    \end{itemize}
    Several algorithms need to be provisioned, among which the most important deals with {intent compilation} for deriving an intent implementation and transitioning an intent to the compiled state.

    \begin{figure}[h!]
        \centering
        \begin{minipage}{0.45\columnwidth}
            \centering
            \includegraphics[width=\textwidth]{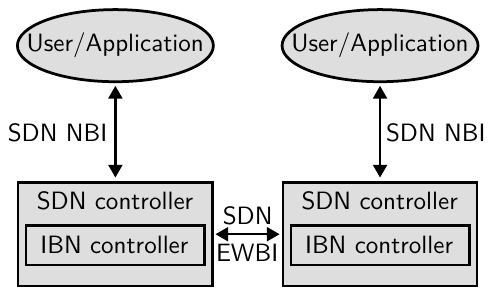}
            \centerline{(a) \ac{IBN}-into-\ac{SDN}}
        \end{minipage}
        \hfill
        \begin{minipage}{0.45\columnwidth}
            \centering
            \includegraphics[width=\textwidth]{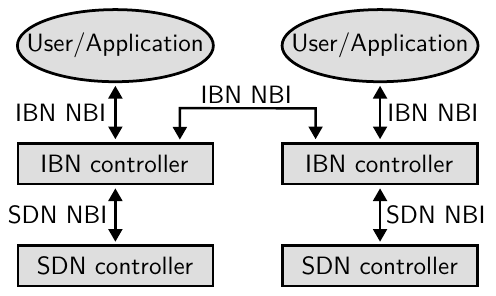}
            \centerline{(b) \ac{IBN}-over-\ac{SDN}}
        \end{minipage}
        \caption{Multi-domain \ac{IBN}/\ac{SDN} architectures.}
        \label{fig:architectures}
    \end{figure}

    When designing a multi-domain \ac{IBN} system, two main architectural approaches exist, as illustrated in Figure~\ref{fig:architectures}.
    First, the \ac{IBN}-into-\ac{SDN} approach integrates the \ac{IBN} framework directly within the \ac{SDN} controller.
    This can lead to complex, monolithic applications where intent handling and device control are intertwined.
    Alternatively, the \ac{IBN}-over-\ac{SDN} architecture used here decouples the two paradigms.
    It separates intent handling into a dedicated framework that operates independently and uses the underlying \ac{SDN} controller strictly as a device driver for physical deployment.
    This separation of concerns fosters modularity and allows each layer to evolve independently.

    \begin{figure}[b]
    \centerline{\includegraphics[width=5cm]{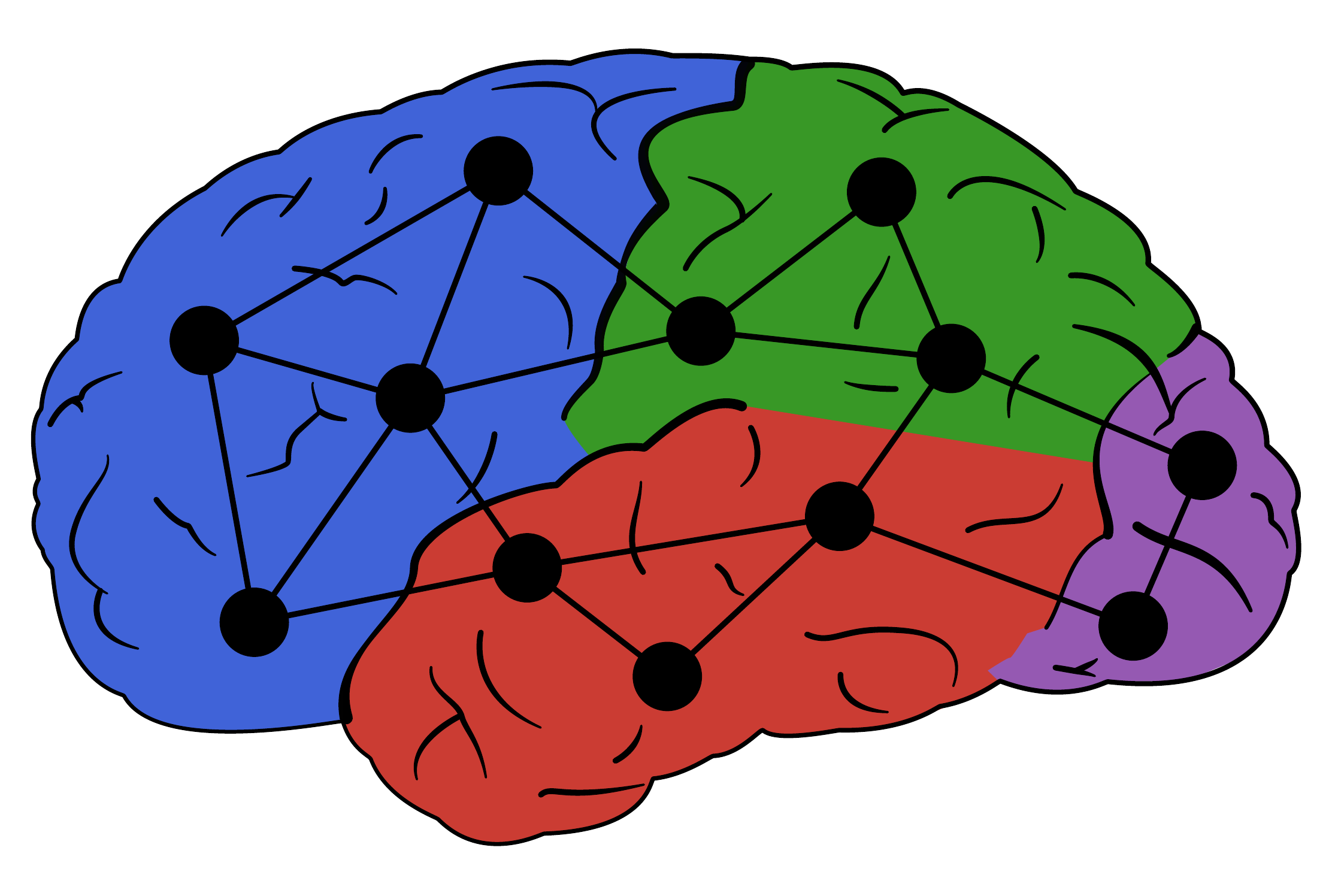}}
    \caption{MINDFul.jl logo}
        \label{fig:logo}
    \end{figure}

\section{MINDFul.jl}
\verb|MINDFul.jl| is an open-source Julia library that provides a playground for \ac{MD} \ac{IBN}.
Although a minimal \ac{MD} \ac{IBN} framework is already implemented, users can extend its functionality.
The library uses intent \acp{DAG} \cite{2023ChristouITG} to represent relationships between different intents.
This powerful scheme connects higher-level intents, which have a logical objective, with low-level intents, which are responsible for resource allocation, using a \ac{DAG}.
This hierarchical intent structure enables seamless interoperability between domains through \emph{intent delegation}, in which an intent node is passed to another domain.

\verb|MINDFul.jl| embraces modularity through the aforementioned \ac{IBN}-over-\ac{SDN} architecture, separating the intent handling from the underlying \ac{SDN} controller.
This ensures that the \ac{IBN} framework focuses solely on intent operations, while the \ac{SDN} controller acts as the device driver.
The library models IP-optical \ac{RMSA} operations in which high-level connectivity intents are recursively compiled into lower-level intents, down to router and \ac{OXC} configurations. 
If an intent spans multiple domains, it is split at border nodes, and the external portion is securely delegated to the adjacent domain without exposing internal topological details.
To evaluate different designs and algorithms under diverse scenarios, the appropriate interfaces are provided to facilitate simulations. 
The library provides a state representation of a common IP-optical network and an \ac{API} for accessing or modifying it. 
The user can use these interfaces to conduct (event-based) simulations.
A companion package, \verb|MINDFulMakie.jl|, can be used for some out-of-the-box visualizations like drawing an intent \ac{DAG} or visualizing a compiled connectivity intent in the network topology.

The architecture provided by \verb|MINDFul.jl| is influenced by previous techno-economical overviews of the equipment used in multilayer metro/core networks \cite{2013Rambach}.
More specifically, we adhere to the pure IP-optical architecture characterized by two layers: the electrical layer, composed of IP routers and virtual links, and the optical layer, composed of \acp{OXC} and physical fiber links.
Given the current technological advancements in coherent pluggable transceivers, such as the OpenZR+ \ac{MSA} \cite{OpenZRProps}, we also seek to incorporate new trends into the package's architecture.

\section{Conclusion}
\verb|MINDFul.jl| is a Julia open-source library that facilitates research on \ac{MD} intent-driven IP-optical networks.
It provides a way to develop experimental intent system architectures and customized algorithms, as well as evaluate them using simulations and visualizations.
We believe that the modular architecture, together with Julia's attributes such as speed and dynamicity, can significantly advance the adaptation of \ac{IBN} in modern \ac{MD} networks.

% **************GENERATED FILE, DO NOT EDIT**************

\bibliographystyle{juliacon}
\bibliography{ref.bib}

\end{document}